*IET-UK International Conference on Information and Communication Technology in Electrical Sciences (ICTES 2007),
Dr. M.G.R. University, Chennai, Tamil Nadu, India. Dec. 20-22, 2007. pp.785-788.*# INTERFERENCE BETWEEN FM CELL SITES AND CDMA CELL SITES

**Puneet Kumar**

S.R.M Engineering College, Chennai, India**Abstract**

Interference is the major problem now days in telecommunication sector. One type of interference which is very common now days is FM Cell sites interference between CDMA Cell sites. Which are the types of interference and various observations during this interference is discussed below in this paper.## 1 Introduction

For reference, a guard zone analysis of the mutual interference between FM cell sites and CDMA mobiles is presented below. The mutual interference between CDMA cell sites and FM mobiles is considered as a, "Guard Band, Guard Zone and Receiver Overload between CDMA and FM Systems". Guard zone recommendations are based on the latter, as it is the dominant interference type.

## 2 Omni Configuration

FM Cell Sites Interfering with CDMA Mobiles. The layout of the interfering FM cells is given below.

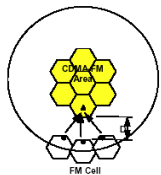

**Figure2.1:** Illustration of the Interfering FM Cell Sites Layout

"FM Cell Sites Interfering with CDMA Mobiles". Figure A-1 depicts the model used to analyze the interference from surrounding FM cell sites affecting CDMA mobiles.

1. Case in which a CDMA mobile has the shortest distance to the surround cells. Ending FM cells.
2. FM channels per cell share the CDMA frequency band. The channels used by all.
3. Surrounding FM cell sites cause interference to the signal reception of the CDMA mobile This interference reduces the CDMA mobile receive *Eb/No*(signal to noise ratio); however, Forward Power Control adjusts the cell traffic channel transmit power in order to Maintain a target Frame Error.
4. This procedure is designed to maintain the mobile receive *Eb/No* at an acceptable level.
5. The FM power falling on the CDMA mobile receive band is attenuated only by propagation loss.
6. Power rises depends upon the distance between the CDMA mobile and the FM cells. If the FM cell sites are located in the proximity of the CDMA mobile, then the required CDMA cell site transmit power per traffic channel may be higher than the maximum allowable value due to excessive FM interference.
7. This situation is obviated by specifying a sufficient buffer area (guard zone) around the CDMA-FM overlay area in which the FM systems can not use the same band as the CDMA system.
8. The required guard zone (*D* in Figure A-1) depends on the Effective Radiated Power (ERP), the FM cell radius *r* and the CDMA cell radius *R*.

Figure A-1. A model for analyzing surrounding FM cell site.

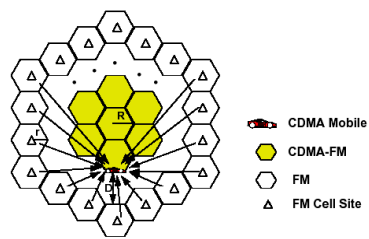

Figure A-1. A model for analyzing surrounding FM cell site interference to CDMA mobiles in a CDMA-FM overlay area

The CDMA cell site traffic channel transmit power required to achieve target*Eb/No* at the mobile is: (Equation A-1)

$$\left(\frac{C}{I}\right)_C = \frac{P_{FMMT} L_{path}(r)}{5 P_{FMMT} L_{path}(d) + I_{CDMAM}(D,R) W_{FM}/W_{CDMA}} \quad \text{(Equation A-2)}$$

Where :( MW) = CDMA cell site traffic channel transmit ERP at 9600 bps

= average propagation loss

= gain of CDMA mobile antenna minus cable loss

= processing gain

785Authorized licensed use limited to: Santa Clara University. Downloaded on July 08,2025 at 09:13:19 UTC from IEEE Xplore. Restrictions apply.

(MW) = CDMA cell site total transmit ERP (average power)
= channel activity factor
(MW) = interference from other CDMA cell sites
(MW) = interference from FM cell sites that is a function of *D*
And *r* (MW) = CDMA mobile receiver noise floor.

In deriving the separation between CDMA-FM mixed cell and FM cell coverage

Needed to maintain the CDMA cell coverage (Figure A-2), the following

Assumptions are made:
- Processing gain = 128
- CDMA mobile receive $E_b/N_o$ = 5.5 dB
- CDMA cell site total average transmit power at the equipment output port

(J4) = 25 W
- Gain of CDMA base antenna minus cable loss = 9 dB
- Gain of CDMA mobile antenna minus cable loss = 0 dB
- CDMA mobile receiver noise floor = -105 dBm (noise figure = 8 dB)
- Channel activity factor = 0.6
- FM cell ERP per channel = 100 W
- Cell site antenna height = 150 ft.
- Mobile antenna height = 5 ft.
- Omni CDMA and FM cells.

To compute the maximum CDMA cell site traffic channel transmit power at 9600 Bps referred to the J4 port, the following data are used:
- the pilot channel power equal to 16% of the total transmit power
- the pilot channel transmit power of 4 W corresponding to a nominal pilot Digital gain of 108
- the traffic channel digital gain between 40 and 80.

The maximum traffic channel transmit power at 9600 bps is given by:

$$4 \times \left(\frac{80}{108}\right)^2 = 2.2\ W = 33.4\ dBm$$

## 3 Surrounding FM Cell Sites Interference to CDMA Mobile in CDMA-FM Overlay Area

1. The conservative threshold for the maximum traffic channel transmits power is chosen as 32 dBm. Since the channel activity factor is 0.6, the maximum average.
2. Power of the forward traffic channel is 1.6 W that is 6.3% of the total transmits Power. Figure A-2 shows the CDMA cell traffic channel transmit power, referred to J4port, required to achieve 5.5 dB of mobile receive $E_b/N_o$ at the cell edge versus

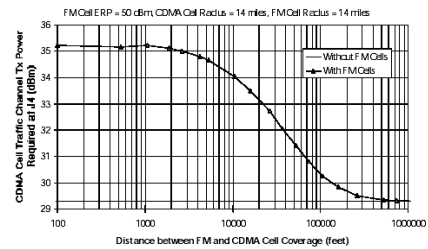

Figure A-2. CDMA cell transmit power required to achieve 5.5 dB $E_b/N_o$ in CDMA mobile at the edge of a boundary cell

3. The separation between CDMA boundary cell and FM cell coverage. The FM and CDMA cell radii are 14 miles. Nominally, the maximum CDMA cell site transmit power per traffic channel is 32 dBm at the J4 port. The figure indicates that, in order to maintain 5.5 dB of CDMA mobile at the cell boundary, the minimum.
4. Distance between the CDMA-FM overlay area and FM cell coverage should be 7 miles. This distance is much less than one tier of FM cells.
5. The corresponding minimum separation required between CDMA-FM mixed cell and FM cell coverage depends upon cell configurations. Three scenarios are Considered:
    - Scenario 1: FM cell radius equals the CDMA cell radius
    - Scenario 2: FM cell radius equals a half of the CDMA cell radius
    - Scenario 3: FM cell radius equals twice the CDMA cell radius.

As the CDMA cell radius increases, the minimum distance between both systems also Increases since the CDMA mobile receives lower desired signal power from its Cell site.

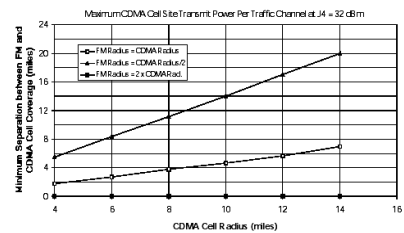

Figure A-3. Separation for preventing FM cell sites from interfering with CDMA mobiles

6. CDMA cell radius is larger than that required for FM cell radius equal to CDMA cell Radius because in the former scenario there are more surrounding FM cell sites with lower propagation loss to the CDMA mobile. Moreover, a guard zone is not necessary when the FM cell radius is equal to twice the CDMA cell radius because this scenario has a smaller number of surrounding FM cell sites with. Figure A-3 shows that all minimum separations required for the suppression of in band Interference from FM cell sites in order to maintain CDMA cell coverage are less than one tier of FM cell except when the FM cell radius equals one half of the CDMA cell radius. If the FM cell site transmits ERP is adjusted based on the cell radius, then the minimum distances required for Scenarios 2 and 3 becomes close.

## 4 CDMA Mobiles Interfering with FM Cell Sites 1

1. The interference power received by the FM cell sites depends on the CDMA mobile transmitted power required for 7 dB of CDMA cell site received *Eb/No*. The transmitted power of a CDMA mobile is a function of CDMA mobile location and CDMA cell site received interference from FM mobiles. The FM cell receive Carrier-to-Interference Ratio (C/I) would be reduced by the in-band interference from the CDMA mobiles.

2. The CDMA mobile carriers falling on the FM cell receive channel is attenuated only by propagation loss. Accordingly, the interference to the FM cell sites depends upon the distance (loss) between the sites and the CDMA mobiles. If the CDMA mobiles are located in the proximity of the FM cell site, then the FM cell receive C/I may be degraded significantly. This situation is obviated by specifying a buffer area (guard zone) around the CDMA-FM overlay area within which the FM systems can not use the same band as CDMA systems. The guard zone (*D* in Figure A-4) depends on the CDMA mobile ERP, the FM cell radius *r* and the CDMA cell radius *R*.

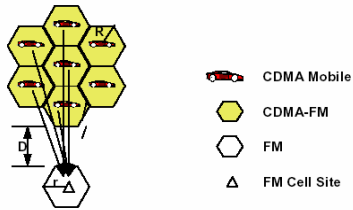

Figure A-4. A model for analyzing CDMA mobile interference to surrounding FM cell sites

4. Since the FM systems in the guard zone and the CDMA systems use different frequency band from the surrounding FM systems, the surrounding FM systems receive interference from five tier-1 analog co-channel cells instead of six. Taking into account the interference generated by the five FM co-channels and by the CDMA mobiles in the 7 CDMA-FM mixed cells, the FM cell receive C/I can be expressed as: (Equation A-2)

$$\left(\frac{C}{I}\right)_C = \frac{P_{FMMT} L_{path}(r)}{5 P_{FMMT} L_{path}(d) + I_{CDMAM}(D,R) W_{FM} / W_{CDMA}}$$ (Equation A-2)

$$\overline{I}_{CDMA}(\overline{d}) = \frac{2\alpha N P_{CR}}{R^2}\left[2\overline{d}^2 \ln\left(\frac{\overline{d}^2}{\overline{d}^2 - R^2}\right) - \frac{R^2(4\overline{d}^4 - 6R^2\overline{d}^2 + R^4)}{2(\overline{d}^2 - R^2)^2}\right]$$ (Equation A-3)

- Number of CDMA users = 20
- FM frequency reuse pattern = 7.

The FM cell C/I degradation due to interference from CDMA mobiles when FM and CDMA cell radii are 14 miles is plotted in Figure A-5. Because the FM frequency reuse pattern is 7, the analog cells receive C/I is 18 dB. FM cell C/I of 17dB is chosen as the performance criterion.

5. When the CDMA cell radius (= FM cell radius) is greater than 1 mile, the FM cell C/I is higher than 18 dB because one FM co-channel interference is replaced with the interference from the CDMA mobiles with lower power.

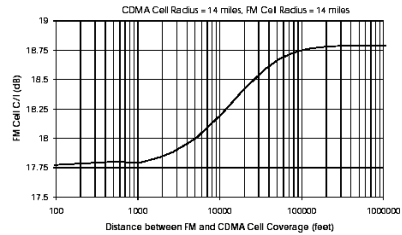

Figure A-5. FM cell C/I degradation due to CDMA mobiles when CDMA cell radius = 14 miles and FM cell radius = 14 miles

7. The minimum separation between CDMA-FM mixed cell and FM cell coverage depends upon cell configuration. Three scenarios are considered:
   - Scenario 1: FM cell radius equals the CDMA cell radius
   - Scenario 2: FM cell radius equals one half of the CDMA cell radius
   - Scenario 3: FM cell radius equals twice the CDMA cell radius.

8. The large CDMA cell radius causes CDMA mobiles to transmit at high power in order to ensure adequate *Eb/No* at their serving site, and the small FM radius results in lower path loss to the FM cell site.

9. Figure A-6 indicates that the interference generated by CDMA mobiles does not play an important role in determining the guard zone between CDMA-FM overlay cells and surrounding FM cells.

"Interference from FM cell sites affecting CDMA MOBILES

1. The separation between CDMA and surrounding FM cells required to prevent FM cell sites from interfering with CDMA mobiles is obtained based on the following

   ASSUMPTIONS: - 3-sector CDMA and FM cells
   - Cell site antenna gain minus cable loss = 12 dB
   - 2 interfering FM channels per sector
   - Processing gain = 128
   - CDMA mobile receive *Eb/No* = 5.5 dB
   - CDMA cell site total transmit power at RF transmit filter output port (J4) = 25 W
   - Gain of CDMA mobile antenna minus cable loss=0 dB

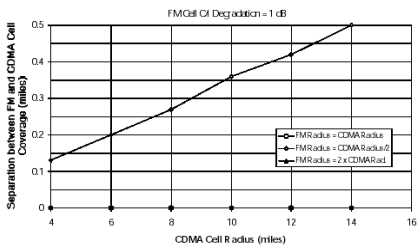

Figure A-6. Separation for preventing CDMA mobiles from interfering with FM cell sites

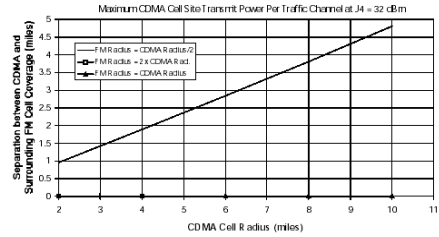

Figure A-7. Separation Required to Prevent FM Cell Sites from Interfering with CDMA Mobiles

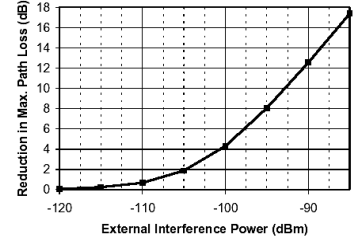

Figure A-8. External Interference Impact on Forward Link Ce Coverage

- CDMA mobile receiver noise floor = -105 dBm (noise figure = 8 dB)
- Channel activity factor = 0.6
- FM cell ERP per channel = 100 W
- Cell site antenna height = 150 ft.
- Mobile antenna height = 5 ft.

3. The minimum separation required versus CDMA cell radius is illustrated in Figure A-7. It is shown that the separation required for the suppression of in-band interference from FM cell sites in order to maintain CDMA cell coverage is one tie of FM cell when the FM cell radius equals a half of the CDMA cell radius. If the FM cell radius is equal to or twice the CDMA cell radius, a separation is not necessary. The separation required for the 3-sector configuration is smaller than that for the omni configuration.

Actually, a CDMA mobile may receive in-band interference from AMPS cell sites but as well as other wireless systems. In the forward link budget examples shown in "Appendix D: Guidelines for 13-kbps Vocoder Option" the Table D-3 and Table D-4, there is no margin allocated for external interference. External interference received by the CDMA mobile causes a penalty in the forward link maximum path loss. In the presence of external interference.

The equation indicates that the penalty in the maximum path loss depends only on the external interference and CDMA mobile receiver noise floor. Figure A-8 shows the reduction in the forward link maximum path loss versus the CDMA mobile received external interference. If external interference is less than -120dBm, then the degradation in the cell coverage is negligible. Service providers can determine a tolerable forward link external interference power level based on the acceptable reduction in the maximum path loss

Interference from CDMA mobiles affecting FM cell sites

To determine the minimum separation between CDMA-FM mixed cell and surrounding FM cell coverage needed to achieve an acceptable FM cell site received C/I, the following data are used:
- 3-sector CDMA and FM cells
- cell site antenna gain minus cable loss = 12 dB
- number of tier-1 analog FM co-channel cells = 1
- FM frequency reuse pattern = 7
- acceptable FM cell site C/I = 17 dB.

Consider a case in which CDMA mobiles are uniformed distributed. Simulations are performed to derive the CDMA mobile transmitted power based on the following assumptions:
- One FM interfering mobile per surrounding analog cell with the shortest distance to the mixed area
- Target CDMA cell site received $E_b/N_o$ = 7 dB
- CDMA cell site receiver noise floor = -107 dBm
- Number of CDMA user/sector = 20
- Interference from CDMA mobiles in other cells = 0.45 times Interference from CDMA mobiles in the serving cell. The separation required versus CDMA cell radius is illustrated in Figure A-9. It is shown that the minimum separation required to mitigate the interference from CDMA mobiles to FM cell sites is one tier of FM cell when the FM cell radius is equal to, twice, or one half of the CDMA cell radius. Unlike the previous three interference types, the separation required for the 3-sector configuration is larger than that for the omni configuration.

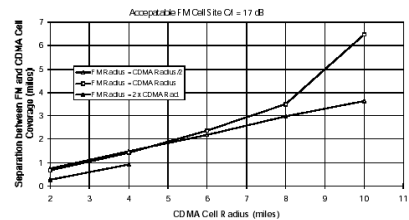

Figure A-9. Separation Required to Prevent CDMA Mobiles from Interfering with FM Cell Sites

## Refrences